\documentclass[10pt]{revtex4}
\usepackage{amssymb}
\usepackage{latexsym}
\usepackage{epsfig}
\usepackage{float}
\usepackage{graphicx,epsfig, color }
\usepackage{graphicx}
\usepackage{subfigure}
\usepackage{amsmath}

\begin{document}
\title{\textbf{Quasi-Topological Magnetic Brane coupled to Nonlinear Electrodynamics}}
\author{A. Bazrafshan$^{1}$ \footnote{Corresponding author}, F. Naeimipour$^{2}$, M. Ghanaatian$^{2}$, Gh. Forozani$^{2}$, A. Alizadeh$^{2}$, }
\address{$^1$ Department of Physics, Jahrom University, 74137-66171 Jahrom, Iran\\$^2$ Department of Physics, Payame Noor University (PNU), P.O. Box 19395-3697 Tehran, Iran
}

\begin{abstract}
In this paper, we are eager to construct a new class of $(n+1)$-dimensional static magnetic brane solutions in quasi-topological gravity coupled to nonlinear electrodynamics such as exponential and logarithmic forms. The solutions of this magnetic brane are horizonless and have no curvature. For $\rho$ near $r_{+}$, the solution $f(\rho)$ is dependent to the values of parameters $q$ and $n$ and for larger $\rho$, it depends on the coefficients of Love-Lock and quasi-topological gravities $\lambda$, $\mu$ and $c$. The obtained solutions also have a conic singularity at $r=0$ with a deficit angle that is only dependent to the parameters $q$, $n$ and $\beta$. We should remind that the two forms of nonlinear electrodynamics theory have similar behaviors on the obtained solutions. At last, by using the counterterm method, we obtain conserved quantities such as mass and electric charge. The value of the electric charge for this static magnetic brane is obtained zero.

\end{abstract}



\maketitle

\section{Introduction}
Modified theories of gravity and nonlinear electrodynamics theory, both are marvelous subjects that can solve many problems.
Models of modified gravities can find a way to unification of the early-time inflation \cite{Star} and late-time cosmic speed-up \cite{Carro,Fay} and also they can make a natural gravitational alternative to dark energy. These models also describe four cosmological phases \cite{Noj1,Noj2}, the hierarchy
problem and the unification of GUTs with gravity which are theories in high energy physics \cite{Noj3}. $f(R)$ gravity is one model of modified gravities in which we can find the galactic dynamics of massive test
particles without the need of dark matter \cite{Capo,Boro}. In an other kind of modified gravities which violation of
the equivalence principle is obvious, the matter Lagrangian density is coupled to an arbitrary function of scalar curvature $R$ \cite{Bert}. An other new kind of modified gravities is quasi-topological gravity which is similar to Lovelock theory with more benefits. As Einstein’s
equations are not the most complete ones in higher
dimensions (n > 4) and they can not satisfy Einstein’s assumptions \cite{Dehg1,Kov}, so, quasi-topological gravity is a higher derivative theory that can solve these problems. This gravity consists of cubic and quartic terms of Riemann tensor and has no limitation on dimensions higher than five, because its terms are not true topological invariants. As this gravity yields to at most second order field equations, this results that the quantization of linearized quasi-topological theory will be free of ghosts.\\
In the other side, nonlinear electrodynamics theory was appeared with the motivations of removing infinite self energy of point-like charges, describing complex systems and chaotic phenomena and behaviors of the compact astrophysical objects such as neutron stars and pulsars, compatibility with AdS/CFT correspondence and string theory frames and description of pair creation for Hawking radiation\cite{Aros,Chen,Fuk}. Nonlinear electrodynamics theory has also been used in applications in cosmological models \cite{Dyad}, such as description of the inflationary epoch and the late-time accelerated expansion of the universe \cite{Nove}. This theory has also been successful to find the first exact regular black hole solutions with a nonlinear electrodynamic source satisfying the weak energy condition \cite{Ayon}. Nonlinear electrodynamics theory has different types which Born-Infeld type is the first one. Born-Infeld form is so important because it naturally comes in the low energy limit of the open string theory and has applications to the description of D-branes and AdS/CFT correspondence \cite{Fradkin}. We can name exponential \cite{Hendi1} and logarithmic \cite{Soleng} Lagrangians as the other types of nonlinear electrodynamics theory which are defined as
\begin{eqnarray}
\mathcal{L}(F)=\left\{
\begin{array}{ll}
$$4\beta^2[\mathrm{exp}(-\frac{F}{4\beta^2})-1]$$,\quad\quad\quad \quad\quad  \ {EN}\quad &  \\ \\
$$-8\beta^2 \mathrm{ln}[1+\frac{F}{8\beta^2}]$$.\quad\quad\quad\quad\quad\quad  \ {LN}\quad &
\end{array}
\right.
\end{eqnarray}
$\beta$ is the nonlinear parameter with dimension
of mass and $F=F_{\mu\nu}F^{\mu\nu}$, where $F_{\mu\nu}$ is the electromagnetic field tensor that is determined as $F_{\mu\nu}=\partial_{\mu}A^{\nu}-\partial_{\nu}A^{\mu}$ and $A_{\mu}$ is the vector potential. We should note that these two Lagrangians reduce to the linear Maxwell Lagrangian as $\beta\rightarrow\infty$.
Like Born-Infeld nonlinear electrodynamics, logarithmic form can remove the infinity of the electric field at the origin \cite{Sheykhi1}, however the exponential
form can not cancel this infinity but it causes a weaker singularity than the one in Einstein-Maxwell theory\cite{Sheykhi2}.\\
The idea of using modified gravities with nonlinear electrodynamics theory can be so interesting which has been studied in many papers. For example, $f(R)$ gravity in the presence of nonlinear electrodynamics has been successful to describe the phenomenon such as power-law inflation and late-time cosmic accelerated expansion due to breaking conformal invariance of the electromagnetic field through a non-minimal gravitational coupling \cite{Bamba}. Similar investigation has been also done in \cite{Hol} which the conformal invariance is not broken. Dilaton black holes and dilaton black branes with nonlinear electrodynamics in four and higher dimensions have been studied in \cite{Sheykhi1,Sheykhi2}. Topological and AdS black holes in Lovelock-Born-Infeld gravity are also in \cite{Aline1,Hend2}. Third order Lovelock black brane in the presence of a nonlinear electromagnetic field has been also investigated in \cite{Pana1}. Magnetic brane solutions of Lovelock gravity with nonlinear electrodynamics have been obtained in \cite{Pana2}.\\
Recently, quasi-topological gravity in the presence of nonlinear electrodynamics has been studied in many papers. For example, Lifshitz quartic quasi-topological black holes in the presence of Born-Infeld electrodynamics have been studied in \cite{Ghana1}. A review of quartic quasi-topological black holes with the nonlinear electromagnetic Born-Infeld field is also presented in \cite{Ghanaa2}. Some of us have also studied the solutions of cubic quasi-topological magnetic branes in the presence of Maxwell and Born-Infeld electromagnetic field in paper \cite{Taghi}. Magnetic branes are attractive because their solutions are horizonless and have a conical geometry. They are
also everywhere flat except at the location of the line source. Now, we have a decision to take a further step
and study the solutions of $(n+1)$-dimensional magnetic branes with exponential and logarithmic nonlinear electrodynamics in quartic quasi-topological gravity.\\
So, we make the structure of this paper as bellow:\\ In section \ref{solu}, we begin with the metric of a horizonless spacetime and an action including nonlinear electrodynamics and quartic quasi-topological theories. Then, we obtain equations and solutions. In the next section \ref{struc}, we investigate physical structure and behavior of the obtained solutions and in section \ref{conserved}, we obtain conserved quantities using the counterterm method. At last section \ref{result}, we write a brief result of the obtained data from this magnetic brane.\\
\section{General formalism}\label{solu}
To have magnetic solutions with no horizons, we start with a metric
with characteristics $(g_{\rho\rho})^{-1}\propto g_{\phi\phi}$ and $g_{tt}\propto -\rho^2$ instead of $(g_{\rho\rho})^{-1}\propto g_{tt}$ and $g_{\phi\phi}\propto -\rho^2$.
So, the $(n+1)$-dimensional metric of a horizonless spacetime with a magnetic brane interpretation is written
\begin{eqnarray}\label{metr}
ds^2=-\frac{\rho^2}{l^2}dt^2+\frac{d\rho^2}{f(\rho)}+l^2 g(\rho)d \phi^2+\frac{\rho^2}{l^2}dX^2,
\end{eqnarray}
where $l$ is a scale factor that is related to the cosmological constant $\Lambda$. $dX^2=\sum_{i=1}^{n-2}$ is a $(n-2)$-dimensional hypersurface with the form of Euclidean metric in volume $V_{n-2}$. $\rho$ and $\phi$ are respectively radial and angular coordinates that $\phi$ is dimensionless and has the range $0\leq \phi <2\pi$.\\
The $(n+1)$-dimensional action in the presence of quartic quasi-topological gravity and nonlinear electrodynamics theory is
\begin{equation}\label{Act1}
I_{bulk}=\frac{1}{16\pi}\int{d^{n+1}x\sqrt{-g}\big\{-2\Lambda+{\mathcal L}_1+\hat{\lambda} {\mathcal L}_2+\hat{\mu} {\mathcal L}_3+\hat{c}{\mathcal L}_4+\mathcal{L}(F)\big\}},
\end{equation}
where $\Lambda=-n(n-1)/2l^2$ and $g$ is the determinant of the metric \eqref{metr}. Einstein-Hilbert, Gauss-Bonnet, cubic and quartic quasi-topological Lagrangians are respectively defined as
\begin{eqnarray}
{\mathcal L}_1=R
\end{eqnarray}
\begin{eqnarray}
{\mathcal L}_2=R_{abcd}R^{abcd}-4R_{ab}R^{ab}+R^2
\end{eqnarray}
\begin{eqnarray}
{{\mathcal L}_3}&=&
R_a{{}^c{{}_b{{}^d}}}R_c{{}^e{{}_d{{}^f}}}R_e{{}^a{{}_f{{}^b}}}+\frac{1}{(2n-1)(n-3)} \bigg(\frac{3(3n-5)}{8}R_{abcd}R^{abcd}R-3(n-1)R_{abcd}R^{abc}{{}_e}R^{de}\nonumber\\
&&+3(n+1)R_{abcd}R^{ac}R^{bd}+6(n-1)R_a{{}^b}R_b{{}^c}R_{c}{{}^a}-\frac{3(3n-1)}{2}R_a{{}^b}R_b{{}^a}R +\frac{3(n+1)}{8}R^3\bigg),
\end{eqnarray}
\begin{eqnarray}
{\mathcal{L}_4}&=& c_{1}R_{abcd}R^{cdef}R^{hg}{{}_{ef}}R_{hg}{{}^{ab}}+c_{2}R_{abcd}R^{abcd}R_{ef}{{}^{ef}}+c_{3}RR_{ab}R^{ac}R_c{{}^b}+c_{4}(R_{abcd}R^{abcd})^2\nonumber\\
&&+c_{5}R_{ab}R^{ac}R_{cd}R^{db}+c_{6}RR_{abcd}R^{ac}R^{db}+c_{7}R_{abcd}R^{ac}R^{be}R^d{{}_e}+c_{8}R_{abcd}R^{acef}R^b{{}_e}R^d{{}_f}\nonumber\\
&&+c_{9}R_{abcd}R^{ac}R_{ef}R^{bedf}+c_{10}R^4+c_{11}R^2 R_{abcd}R^{abcd}+c_{12}R^2 R_{ab}R^{ab}\nonumber\\
&&+c_{13}R_{abcd}R^{abef}R_{ef}{{}^c{{}_g}}R^{dg}+c_{14}R_{abcd}R^{aecf}R_{gehf}R^{gbhd},
\end{eqnarray}
where
\begin{eqnarray}
&&c_{1}=-(n-1)(n^7-3n^6-29n^5+170n^4-349n^3+348n^2-180n+36)\nonumber\\
&&c_{2}=-4(n-3)(2n^6-20n^5+65n^4-81n^3+13n^2+45n-18)\nonumber\\
&&c_{3}=-64(n-1)(3n^2-8n+3)(n^2-3n+3)\nonumber\\
&&c_{4}=-(n^8-6n^7+12n^6-22n^5+114n^4-345n^3+468n^2-270n+54)\nonumber\\
&&c_{5}=16(n-1)(10n^4-51n^3+93n^2-72n+18)\nonumber\\
&&c_{6}=-32(n-1)^2(n-3)^2(3n^2-8n+3)\nonumber\\
&&c_{7}=64(n-2)(n-1)^2(4n^3-18n^2+27n-9)\nonumber\\
&&c_{8}=-96(n-1)(n-2)(2n^4-7n^3+4n^2+6n-3)\nonumber\\
&&c_{9}=16(n-1)^3(2n^4-26n^3+93n^2-117n+36)\nonumber\\
&&c_{10}=n^5-31n^4+168n^3-360n^2+330n-90\nonumber\\
&&c_{11}=2(6n^6-67n^5+311n^4-742n^3+936n^2-576n+126)\nonumber\\
&&c_{12}=8(7n^5-47n^4+121n^3-141n^2+63n-9)\nonumber\\
&&c_{13}=16n(n-1)(n-2)(n-3)(3n^2-8n+3)\nonumber\\
&&c_{14}=8(n-1)(n^7-4n^6-15n^5+122n^4-287n^3+297n^2-126n+18).\nonumber\\
\end{eqnarray}
$\hat{\lambda}$, $\hat{\mu}$ and $\hat{c}$ are respectively the parameters of Gauss-Bonnet, cubic and quartic quasi-topological Lagrangians
\begin{eqnarray}
\hat{\lambda}=\frac{\lambda L^2}{(n-2)(n-3)},
\end{eqnarray}
\begin{eqnarray}
\hat{\mu}=\frac{8\mu(2n-1)l^4}{(n-2)(n-5)(3n^2-9n+4)},
\end{eqnarray}
\begin{eqnarray}
\hat{c}=\frac{cl^6}{n(n-1)(n-3)(n-7)(n-2)^2(n^5-15n^4+72n^3-156n^2+150n-42)}.
\end{eqnarray}

The magnetic field is associated with the angular component $A_{\phi}$. So, we introduce the gauge potential for the static solutions as
\begin{eqnarray}\label{h1}
A_{\mu}=h(\rho)\delta_{\mu}^{\phi}.
\end{eqnarray}
Using the above relations in action \eqref{Act1} and integrating by parts, we can get to the action
\begin{eqnarray}\label{Act2}
S&=&\frac{n-1}{16\pi l^2}\times\nonumber\\
&&\int d^{n} x\int{d\rho N(\rho)\bigg\{\bigg[\rho^n\bigg(1+\Psi+\lambda\Psi^2+\mu\Psi^3+c\Psi^4\bigg)\bigg]^{'}+
\left\{
\begin{array}{ll}
$$\frac{4\beta^2 l^2 \rho^{n-1}}{n-1}\bigg[\mathrm{exp}\bigg(-\frac{h^{'2}}{2l^2\beta^2 N^2(\rho)}\bigg)-1\bigg]\bigg\}$$,\quad  \ {EN}\quad &  \\ \\
$$-\frac{8\beta^2 l^2 \rho^{n-1}}{n-1}\mathrm {ln}\bigg[1+\frac{h^{'2}}{4\beta^2 l^2N^2(\rho)}\bigg]\bigg\}$$,\quad\quad\quad  \ {LN}\quad &
\end{array}
\right.}
\end{eqnarray}
where $\Psi(\rho)=-\frac{l^2}{\rho^2}f(\rho)$,
$g(\rho)=N^2(\rho)f(\rho)$ and prime shows the first derivative with respect to $\rho$. EN and LN are respectively the abbreviation of Exponential and Logarithmic Nonlinear.
By varying this action with respect to function $\Psi(\rho)$, it leads to the equation
\begin{equation}\label{equ1}
\{1+2\lambda \Psi(\rho)+3\mu \Psi^2(\rho)+4c\Psi^3(\rho)\}N^{'}(\rho)=0.
\end{equation}
The above equation shows that $N(\rho)$ must be a constant value which we choose $N(\rho)=1$. By varying the action \eqref{Act2} with respect to the functions $N(\rho)$ and $h(\rho)$ and using the obtained condition $N(\rho)=1$(or $f(\rho)=g(\rho)$), we get to the equations
\begin{equation}\label{equ2}
\bigg\{(n-1)\rho^n\bigg(1+\Psi+\lambda \Psi^2+\mu \Psi^3+c\Psi^4\bigg)\bigg\}^{'}+\left\{
\begin{array}{ll}
$$4\rho^{n-1}(l^2 \beta^2+h^{'2})\mathrm {exp}\bigg(-\frac{h^{'2}}{2 l^2\beta^2}\bigg)-4l^2\beta^2\rho^{n-1}=0$$,\quad\quad \quad\quad\quad  \ {EN}\quad &  \\ \\
$$-8\beta^2 l^2 \rho^{n-1} \mathrm{ln} (1+\frac{h^{'2}}{4\beta^2 l^2})+4\rho^{n-1}h^{'2}(1+\frac{h^{'2}}{4\beta^2 l^2})^{-1}=0$$,\quad\quad\quad  \ {LN}\quad &
\end{array}
\right.
\end{equation}
and
\begin{equation}\label{equ3}
\left\{
\begin{array}{ll}
$$\bigg(\rho^{n-1}h^{'}\mathrm {exp}\bigg[-\frac{h^{'2}}{2l^2\beta^2}\bigg]\bigg)^{'}=0$$,\quad \quad\quad\quad \quad\quad\quad\quad  \ {EN}\quad &  \\ \\
$$\bigg(\rho^{n-1}h^{'}(1+\frac{h^{'2}}{4\beta^2 l^2})^{-1}\bigg)^{'}=0$$.\quad\quad\quad\quad \quad\quad\quad\quad\ {LN}\quad &
\end{array}
\right.
\end{equation}
If we solve the equation \eqref{equ3}, we get to electromagnetic field
\begin{eqnarray}\label{Fphir}
F_{\phi \rho}=h^{'}=\left\{
\begin{array}{ll}
$$l\beta\sqrt{-L_{W}(-\eta)}$$,\quad \quad\quad\quad \quad\quad\quad\quad\quad\quad  \ {EN}\quad &  \\ \\
$$\frac{2ql^{n-2}}{\rho^{n-1}}(1+\sqrt{1-\eta})^{-1}$$,\quad\quad\quad\quad\quad\quad\quad  \ {LN}\quad &
\end{array}
\right.
\end{eqnarray}
where $\eta=\frac{q^2l^{2n-6}}{\beta^2 \rho^{2n-2}}$ and $q$ is the constant of integration. $L_{W}$ is the Lambert function that has the following series expansion
\begin{eqnarray}
L_{W}(x)=x-x^2+\frac{3}{2}x^3+... .
\end{eqnarray}
If we expand $F_{\phi\rho}$ for large $\beta$, we get to
\begin{eqnarray}
F_{\phi\rho}=\frac{ql^{n-2}}{\rho^{n-1}}+\left\{
\begin{array}{ll}
$$\frac{q^3 l^{3n-8}}{2\beta^2\rho^{3n-3}}+\mathcal{O}(\frac{1}{\beta^4})$$,\quad \quad\quad\quad\quad\quad\quad  \ {EN}\quad &  \\ \\
$$\frac{q^3 l^{3n-8}}{4\beta^2\rho^{3n-3}}+\mathcal{O}(\frac{1}{\beta^4})$$,\quad\quad\quad\quad\quad\quad\quad  \ {LN}\quad &
\end{array}
\right.
\end{eqnarray}
where the first term is the electromagnetic field of magnetic brane in the presence of linear Maxwell theory in higher dimensions \cite{Taghi} and the next terms are the corrections to the electromagnetic field, in the presence of nonlinear electrodynamics. By reminding that the vector potential $A_{\phi}$ is only dependent to coordinate $\rho$, we get to the relation $F_{\phi\rho}=-\partial_{\rho}A_{\phi}$ that by solving $A_{\phi}=-\int{F_{\phi\rho}d\rho}$, we obtain
\begin{eqnarray}\label{A1}
A_{\phi}=\left\{
\begin{array}{ll}
$$-\frac{n-1}{n-2}l\beta\big(\frac{l^{n-3}q}{\beta}\big)^{\frac{1}{n-1}}\big(-L_{W}(-\eta)\big)^{\frac{n-2}{2(n-1)}}\bigg\{ {}_2F_{1}\bigg(\big[\frac{n-2}{2(n-1)}\big]\,,\big[\frac{3n-4}{2(n-1)}\big]\,,-\frac{1}{2(n-1)}L_{W}(-\eta)\bigg)\\
-\frac{n-2}{n-1} \mathrm {exp}(-\frac{1}{2(n-1)}L_{W}(-\eta))\bigg\}$$,\quad\quad\quad\quad\quad\quad\quad\quad\quad\quad\quad\quad\quad\quad\quad\quad\quad\quad\quad\quad\quad\quad\quad\quad\quad  \ {EN}\quad &  \\ \\
$$\frac{ql^{n-2}}{(n-2)\rho^{n-2}}{}_{3}F_{2}([\frac{n-2}{2(n-1)},\frac{1}{2},1]\,,[\frac{3n-4}{2(n-1)},2]\,,\eta)$$.\quad\quad\quad\quad\quad\quad\quad\quad\quad\quad\quad\quad\quad\quad\quad\quad\quad \quad \quad\quad  \ {LN}\quad &
\end{array}
\right.
\end{eqnarray}
As $\beta\rightarrow\infty$, we get to
\begin{eqnarray}
A_{\phi}=\frac{ql^{n-2}}{(n-2)\rho^{n-2}},
\end{eqnarray}
which is the $(n+1)$-dimensional vector potential of Maxwell theory \cite{Taghi}.
Using Eq. \eqref{Fphir} in Eq. \eqref{equ2} leads to the relation
\begin{eqnarray}\label{equ4}
c \Psi^4+\mu \Psi^3+\lambda \Psi^2+\Psi+\kappa=0,
\end{eqnarray}
where $\kappa$ is
\begin{eqnarray}\label{kappa1}
\kappa &=& 1-\frac{M}{(n-1)\rho^n}\nonumber\\
&&+\left\{
\begin{array}{ll}
$$-\frac{4 l^2\beta^2}{n(n-1)}-\frac{4(n-1)\beta q l^{n-1}}{n(n-2)\rho^n}(\frac{l^{n-3}q}{\beta})^{\frac{1}{n-1}}(-L_{W}(-\eta))^{\frac{n-2}{2(n-1)}}\times{}_2 F_{1}([\frac{n-2}{2(n-1)}]\,,[\frac{3n-4}{2(n-1)}]\,,-\frac{1}{2(n-1)}L_{W}(-\eta))\\+\frac{4\beta q l^{n-1}}{(n-1)\rho^{n-1}}[-L_{W}(-\eta)]^{\frac{1}{2}}\times[1+\frac{1}{n}(-L_{W}(-\eta))^{-1}]$$,\quad \quad\quad\quad\quad\quad \quad\quad\quad\quad\quad\quad\quad\quad\quad\quad\quad\quad\quad\quad  \ {EN}\quad &  \\ \\
$$\frac{8(2n-1)}{n^2(n-1)}\beta^2 l^2[1-\sqrt{1-\eta}]-\frac{8(n-1)q^2l^{2n-4}}{n^2 (n-2)\rho^{2n-2}}{}_2 F_{1}([\frac{n-2}{2(n-1)},\frac{1}{2}]\,,[\frac{3n-4}{2(n-1)}]\,,\eta)-\frac{8}{n(n-1)}l^2\beta^2 \mathrm {ln}[\frac{2-2\sqrt{1-\eta}}{\eta}]$$,\quad  \ {LN}\quad &
\end{array}
\right.
\end{eqnarray}
and $M$ is the integration constant and is related to the mass of this magnetic brane. In the above solution, we have used the following relation for the Lambert function
\begin{eqnarray}
L_{W}(x)e^{L_{W}(x)}=x.
\end{eqnarray}
To have real solutions for the equation \eqref{equ4}, the condition
\begin{eqnarray}\label{Delta}
\Delta=\frac{H^2}{4}+\frac{P^3}{27}>0,
\end{eqnarray}
should be satisfied, where $P$ and $H$ are defined as
\begin{eqnarray}
P&=&-\frac{\alpha^2}{12}-\gamma\,\,\,\,\,\,\,\,\,\,,\,\,\,\,\,\,\,
H=-\frac{\alpha^3}{108}+\frac{\alpha \gamma}{3}-\frac{\beta^2}{8},
\end{eqnarray}
that $\alpha$, $\beta$ and $\gamma$ are
\begin{eqnarray}\label{eq1}
\alpha&=&\frac{-3\mu^2}{8c^2}+\frac{\lambda}{c}\,\,\,\,\,\,\,\,\,\,\,,\,\,\,\,\,\,
\beta=\frac{\mu^3}{8c^3}-\frac{\mu\lambda}{2c^2}+\frac{1}{c}\nonumber\\
&&\gamma=\frac{-3\mu^4}{256c^4}+\frac{\lambda \mu^2}{16c^3}-\frac{\mu}{4c^2}+\frac{\kappa}{c}.
\end{eqnarray}
If we define the following definitions,
\begin{eqnarray}
U=\bigg(-\frac{H}{2}\pm\sqrt{\Delta}\bigg)^{\frac{1}{3}},
\end{eqnarray}
\begin{equation}
y=\left\{
\begin{array}{ll}
$$-\frac{5}{6}\alpha+U-\frac{P}{3U}$$,\quad \quad\quad\quad \  \ {U\neq 0,}\quad &  \\ \\
$$-\frac{5}{6}\alpha+U-\sqrt[3]{H}$$, \quad \quad\quad\quad{U=0,}\quad &
\end{array}
\right.
\end{equation}
\begin{eqnarray}
W=\sqrt{\alpha+2y},
\end{eqnarray}
the solution $f(\rho)$ for the Eq.\eqref{equ4} is obtained as
\begin{eqnarray}
f(\rho)=\frac{-\rho^2}{l^2}\bigg(-\frac{\mu}{4c}+\frac{\pm_{s}W\mp_{t}\sqrt{-(3\alpha+2y\pm_{s}\frac{2\beta}{W})}}{2}\bigg).
\end{eqnarray}
In the above equation, two $\pm_{s}$ should have both the same sign, while the sign of $\pm_{t}$ is independent. It is noteworthy
to say that in order to have the cubic quasi-topological or Gauss-Bonnet solutions, we should replace $\mu=0$ or $\lambda=0$ in the equation \eqref{equ4} and find the solutions not in the above relations because we get to the vague values.

\section{Physical properties of the solutions}\label{struc}
\begin{figure}
\centering
\subfigure[EN]{\includegraphics[scale=0.35]{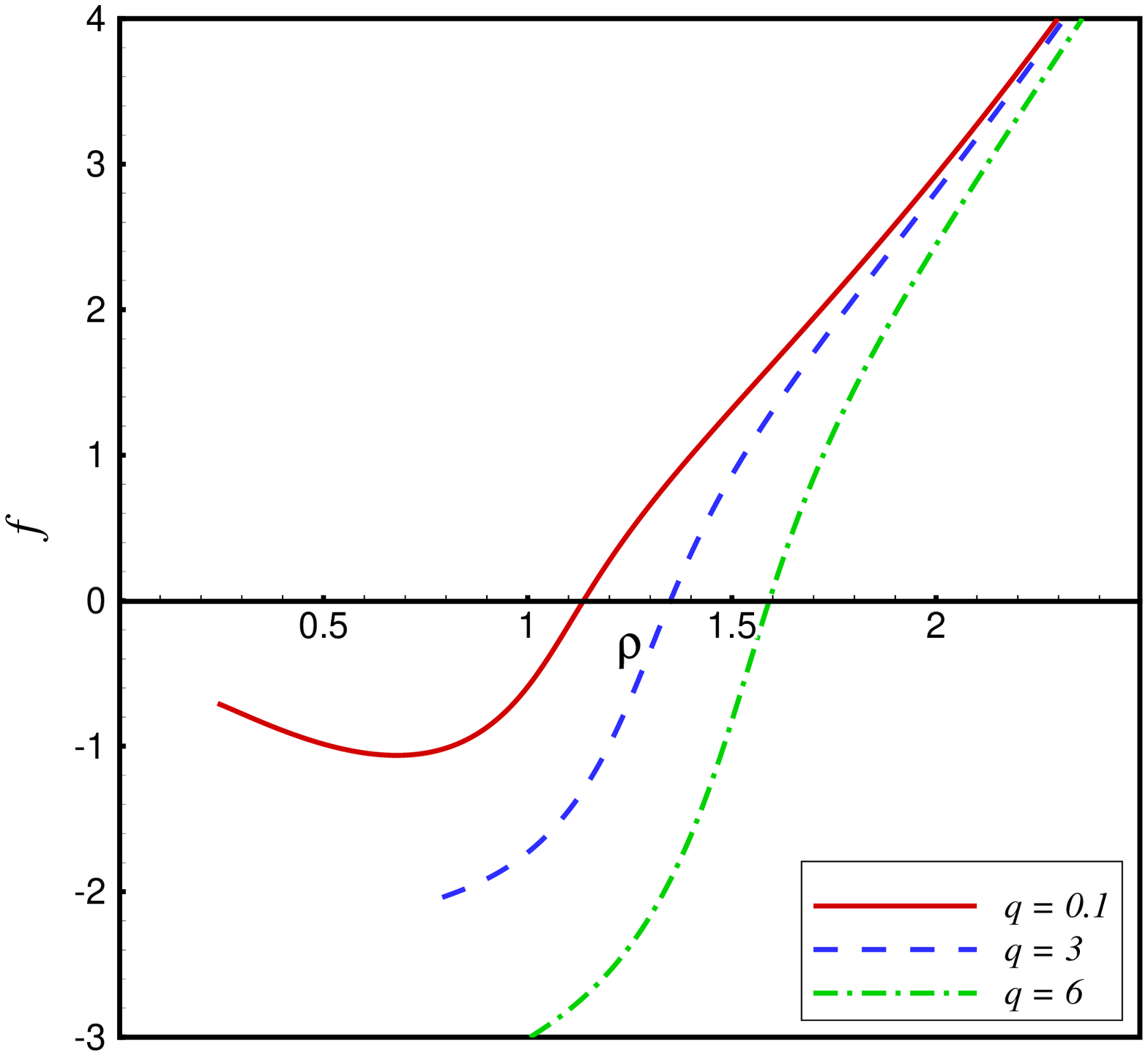}\label{fig1a}}\hspace*{.2cm}
\subfigure[LN]{\includegraphics[scale=0.35]{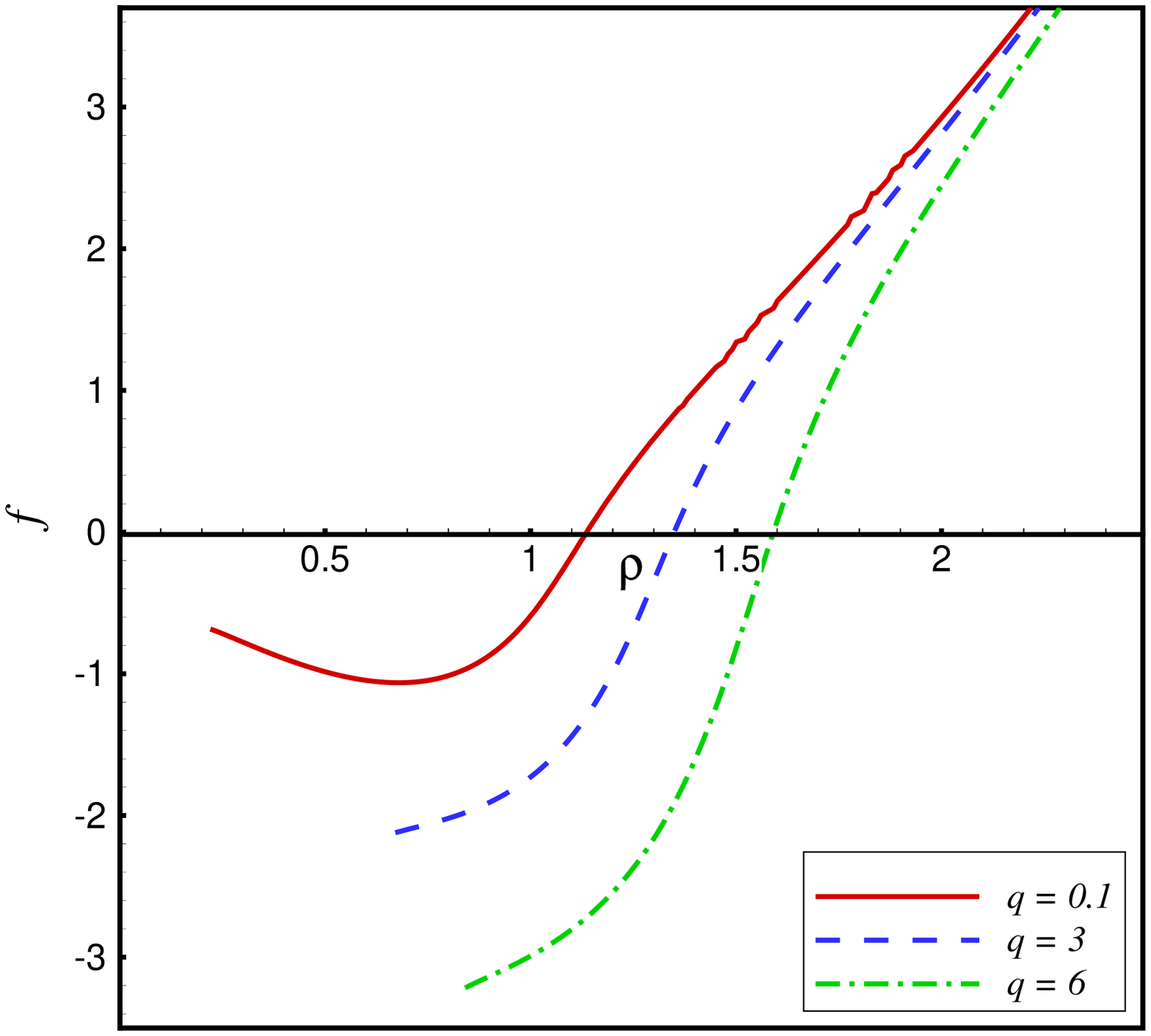}\label{fig1b}}\caption{$f(\rho)$ versus $\rho$ with $M=5$, $\beta=10$, $n=4$, $\lambda=-0.01$, $\mu=0.4$ and $c=-0.01$.}\label{figure1}
\end{figure}

\begin{figure}
\centering
\subfigure[EN]{\includegraphics[scale=0.35]{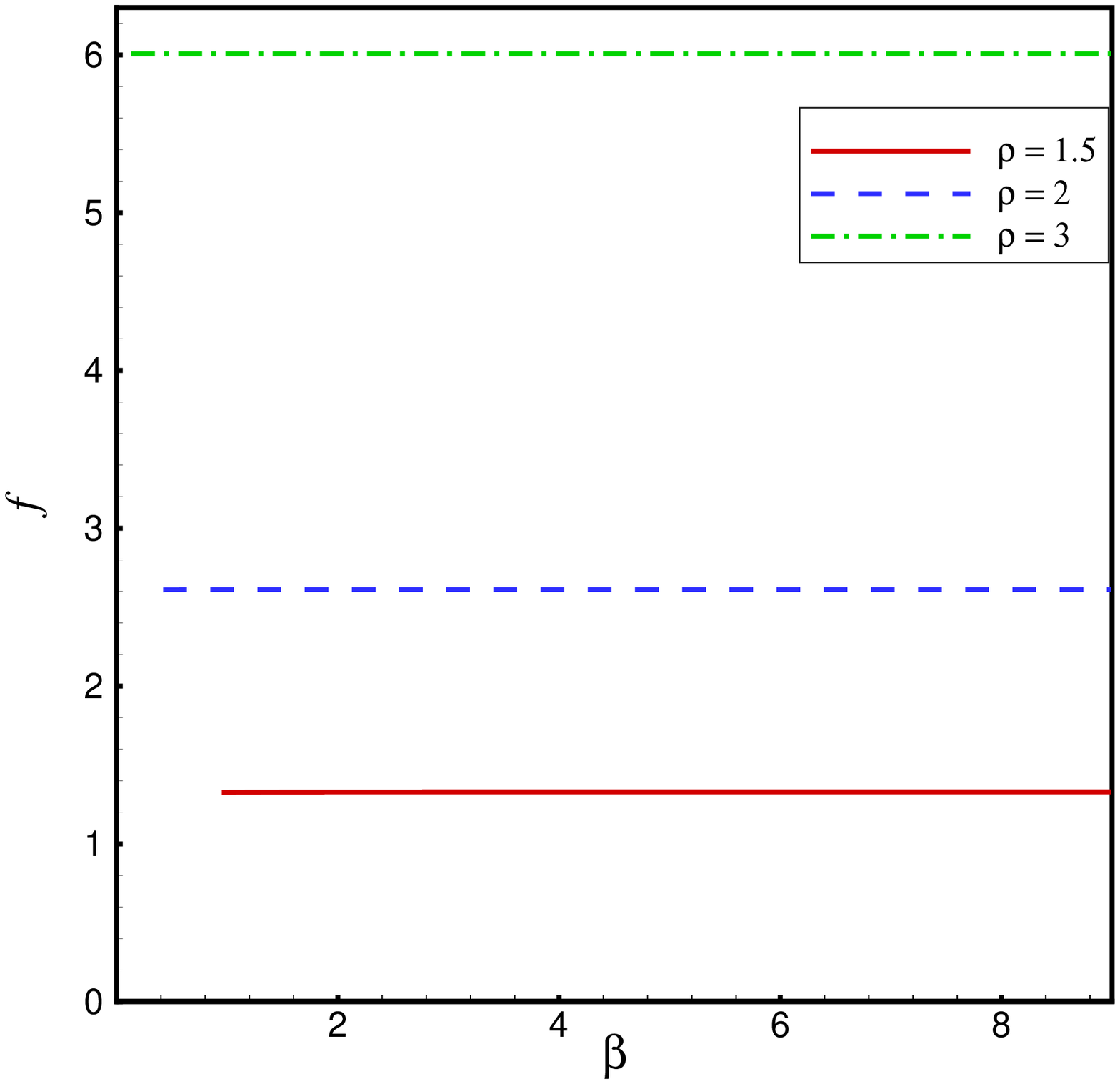}\label{fig2a}}\hspace*{.2cm}
\subfigure[LN]{\includegraphics[scale=0.35]{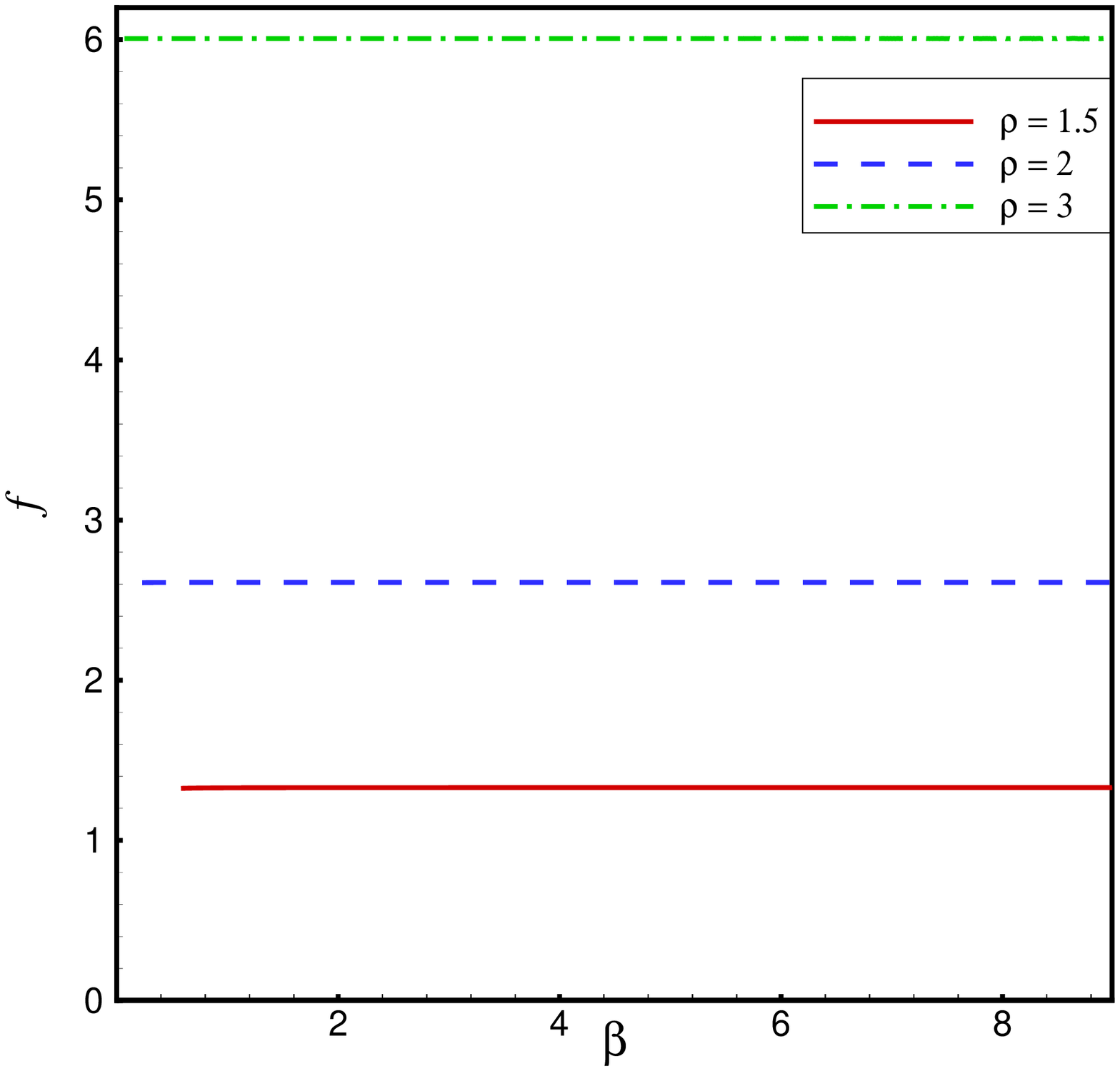}\label{fig2b}}\caption{ $f(\rho)$ versus $\beta$ with $M=1$, $q=2$, $n=4$, $\lambda=0.01$, $\mu=1.1$ and $c=-0.02$.}\label{figure2}
\end{figure}

\begin{figure}
\centering
\subfigure[$M=2$, $q=3$, $\beta=5$, $n=4$, $\mu=-0.4$ and $c=-0.01$]{\includegraphics[scale=0.27]{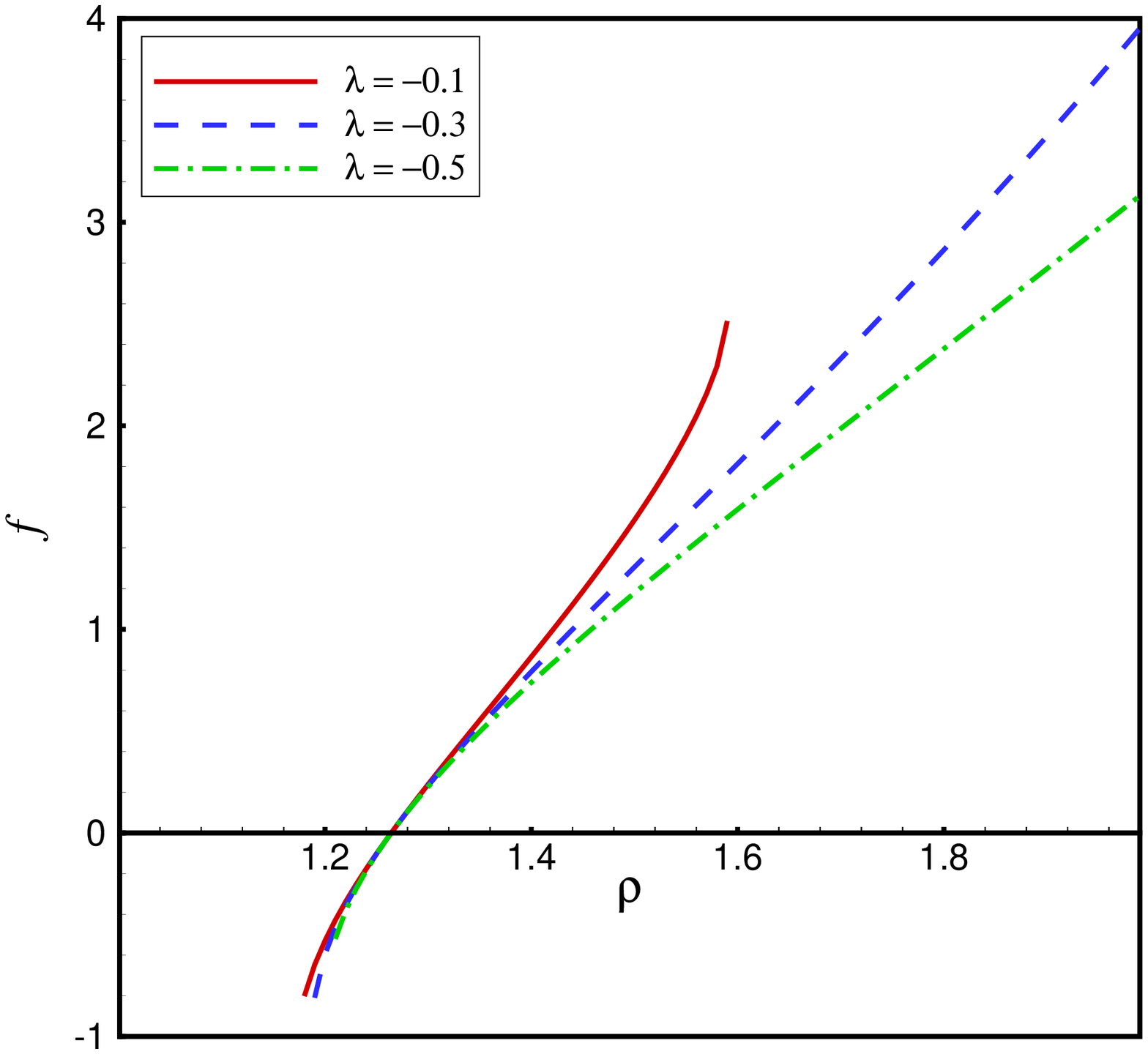}\label{fig3a}}\hspace*{.2cm}
\subfigure[$M=10$, $q=1$, $\beta=2$, $n=4$, $\lambda=-0.01$ and $c=-0.01$]{\includegraphics[scale=0.27]{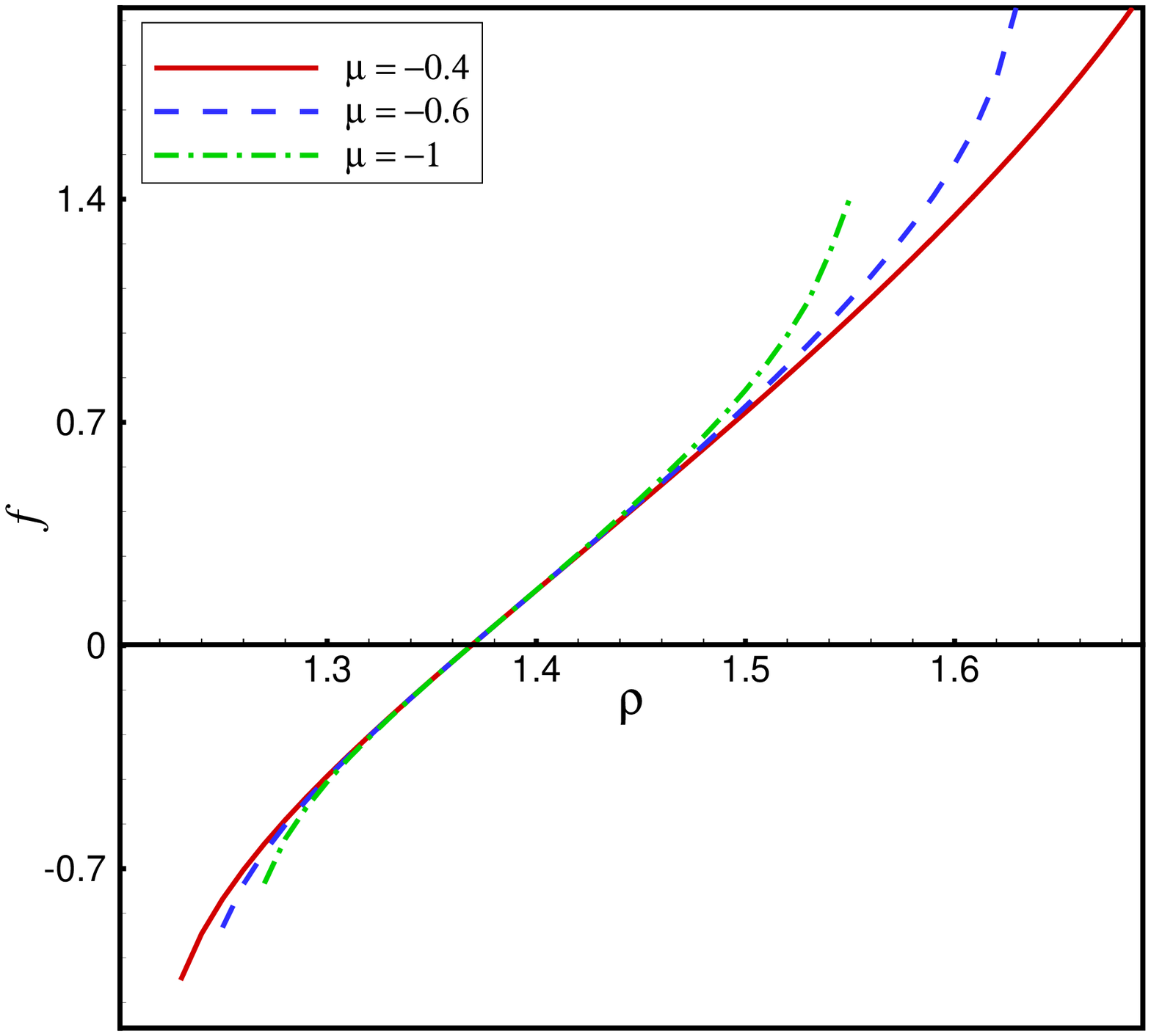}\label{fig3b}}\hspace*{.2cm}
\subfigure[$M=5$, $q=2$, $\beta=5$, $n=4$, $\lambda=-0.2$ and $\mu=-0.5$]{\includegraphics[scale=0.27]{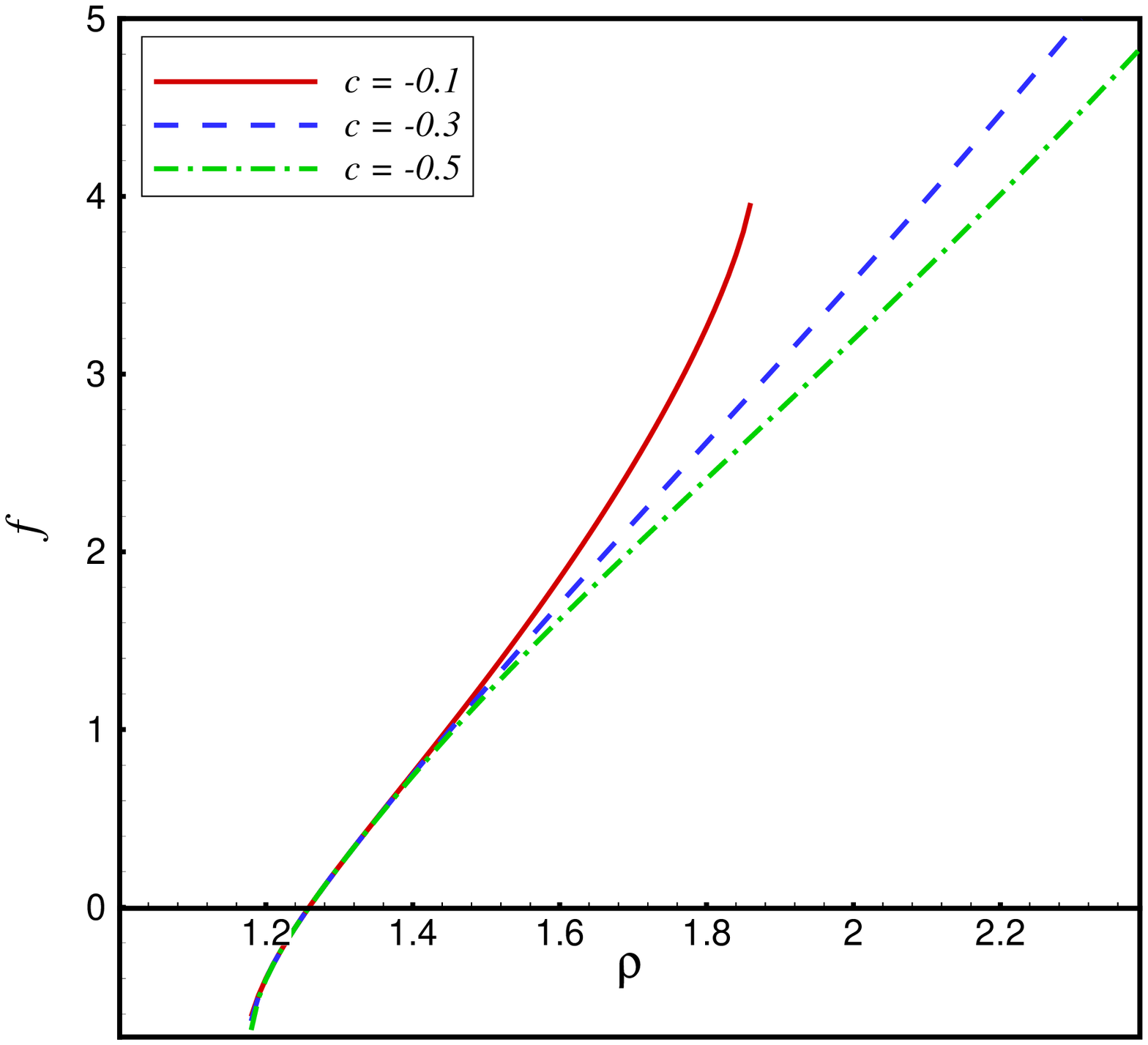}\label{fig3c}}\caption{$f(\rho)$ versus $\rho$ for EN electrodynamics}\label{figure3}
\end{figure}

\begin{figure}
\center
\includegraphics[scale=0.5]{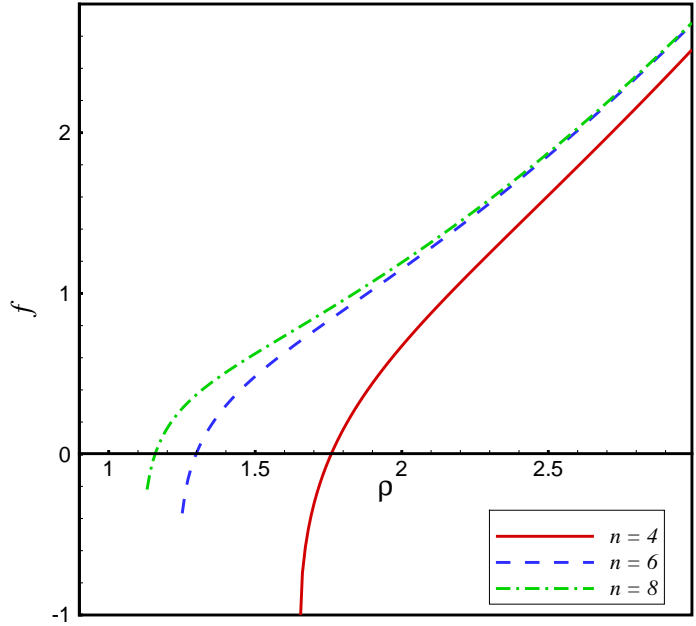}
\caption{\small{$f(\rho)$ versus $\rho$ with $M=5$, $q=3$, $\beta=10$, $\lambda=-0.01$, $\mu=0.4$ and $c=-0.1$ for EN electrodynamics.} \label{figure4}}
\end{figure}

In this section, we have the aim to investigate the geometric and physical properties of the solutions like horizons, singularity and behaviors of the function $f(\rho)$. As we know, to find the horizons of the obtained solutions, the condition $f(r_{+})=0$ should be satisfied where $r_{+}$ is the horizon. Suppose that $r_{+}$ is the largest real root of $f(\rho)=0$, which leads that the function $f(\rho)$ is  positive for $\rho>r_{+}$ and negative for $\rho<r_{+}$. So, in the range of $\rho<r_{+}$, $g_{\rho\rho}$ and followed by $g_{\phi\phi}$ are negative which results that the signature of the metric changes from $(-++++)$ to $(---++)$ that is not acceptable. Therefore, we delete this unacceptable range $0<\rho<r_{+}$ and so the function $f(\rho)$ is limited to the acceptable range $\rho>r_{+}$. For the ease, we can use a suitable transformation as
\begin{eqnarray}\label{trans}
r=\sqrt{\rho^2-r_{+}^2} \Rightarrow d\rho^2=\frac{r^2}{r^2+r_{+}^2}dr^2,
\end{eqnarray}
which results to
\begin{eqnarray}\label{metric2}
ds^2=-\frac{r^2+r_{+}^2}{l^2}dt^2+\frac{r^2 dr^2}{(r^2+r_{+}^2)f(r)}+l^2 g(r)d \phi^2+\frac{r^2+r_{+}^2}{l^2}dX^2.
\end{eqnarray}
By this transformation, $r$ has the range $0\leq r<\infty$ which $f(r)$ is positive and real for $0< r<\infty$ and zero for $r=0$.
This transformation also leads to the changes for $F_{\phi r}$ and $\kappa$ as
\begin{eqnarray}\label{Fphir1}
F_{\phi r}=\left\{
\begin{array}{ll}
$$l\beta\sqrt{-L_{W}(-\eta)}$$,\quad \quad\quad\quad \quad\quad\quad\quad\quad\quad  \ {EN}\quad &  \\ \\
$$\frac{2ql^{n-2}}{(r^2+r_{+}^2)^{\frac{n-1}{2}}}(1+\sqrt{1-\eta})^{-1}$$,\quad\quad\quad\quad\quad \ {LN}\quad &
\end{array}
\right.
\end{eqnarray}
\begin{eqnarray}\label{kappa2}
\kappa&=& 1-\frac{M}{(n-1)(r^2+r_{+}^2)^{\frac{n}{2}}}\nonumber\\
&&+\left\{
\begin{array}{ll}
$$-\frac{4 l^2\beta^2}{n(n-1)}-\frac{4(n-1)\beta q l^{n-1}}{n(n-2)(r^2+r_{+}^2)^{\frac{n}{2}}}(\frac{l^{n-3}q}{\beta})^{\frac{1}{n-1}}(-L_{W}(-\eta))^{\frac{n-2}{2(n-1)}}\times{}_2 F_{1}([\frac{n-2}{2(n-1)}]\,,[\frac{3n-4}{2(n-1)}]\,,\\
-\frac{1}{2(n-1)}L_{W}(-\eta))+\frac{4\beta q l^{n-1}}{(n-1)(r^2+r_{+}^2)^{\frac{n-1}{2}}}[-L_{W}(-\eta)]^{\frac{1}{2}}\times[1+\frac{1}{n}(-L_{W}(-\eta))^{-1}]
$$,\quad\quad\quad\quad  \ {EN}\quad &  \\ \\
$$\frac{8(2n-1)}{n^2(n-1)}\beta^2 l^2[1-\sqrt{1-\eta}]-\frac{8(n-1)q^2l^{2n-4}}{n^2 (n-2)(r^2+r_{+}^2)^{n-1}}{}_2 F_{1}([\frac{n-2}{2(n-1)},\frac{1}{2}]\,,[\frac{3n-4}{2(n-1)}]\,,\eta)\\-\frac{8}{n(n-1)}l^2\beta^2 \mathrm {ln}[\frac{2-2\sqrt{1-\eta}}{\eta}]$$,\quad\quad \quad\quad\quad\quad \quad\quad\quad\quad \quad\quad\quad\quad \quad\quad\quad\quad \quad\quad\quad\quad \quad\quad\quad  \ {LN}\quad &
\end{array}
\right.
\end{eqnarray}
where $\eta=\frac{q^2l^{2n-6}}{\beta^2(r^2+r_{+}^2)^{n-1}}$. To find the singularity of the obtained solutions, we calculate Kretschmann scalar,
\begin{eqnarray}
\mathcal{K}=R_{\mu\nu\alpha\beta}R^{\mu\nu\alpha\beta}=f^{''2}+\frac{2(n-1)}{\rho^2}f^{'2}+\frac{2(n-1)(n-2)}{\rho^4}f^2,
\end{eqnarray}
where double prime is the second derivative of the function $f$ with respect to $\rho$. It seems that the solutions have a singularity at $\rho=0$ because the Kretschmann scalar diverges at this point, but, as we found out, the point $\rho=0$ is not in the acceptable range of $\rho$. So, this magnetic brane has no singularity.\\
To know more about this magnetic brane, we have studied the behavior of $f(\rho)$ in figures \eqref{figure1}-\eqref{figure4}. For this purpose, we have considered $l=1$ without losing the issue. In Fig. \ref{figure1}, we have plotted $f(\rho)$ versus $\rho$ for different values of $q$ and for EN(Exponential Nonlinear) and LN(Logarithmic Nonlinear) electrodynamics. According to our previous sayings, there is a $r_{+}$ which $f(\rho)<0$ for $\rho<r_{+}$ and this part is not acceptable. In Fig. \ref{figure1}, for constant values of parameters $M$, $\beta$, $n$, $\lambda$, $\mu$, $c$, the value of $r_{+}$ depends on the value of $q$ and it increases by increasing the value of $q$. Also, for a definite value of $q$, the value of $r_{+}$ is independent of the kinds of the nonlinear electrodynamics, Exponential or Logarithmic. The function $f$ is not also sensitive to the value of $q$ for large $r_{+}$ and has a constant value for each value of $\rho$, but in the region near $r_{+}$, it decreases as $q$ increases.\\
In Fig. \ref{figure2}, we are eager to know the behavior of $f$ versus $\beta$ for different values of $\rho$. In each Figs. \ref{fig2a} and \ref{fig2b}, for fixed parameters $M$, $q$, $n$, $\lambda$, $\mu$ and $c$, there is a $\beta_{\mathrm{min}}$ for each value of $\rho$ which the function $f$ is not real for $\beta<\beta_{\mathrm{min}}$ and it has a constant value for $\beta>\beta_{\mathrm{min}}$. The value of $\beta_{\mathrm{min}}$ depends on the value of $\rho$ and increases by decreasing the value of $\rho$. It is also clear that Fig. \ref{fig2b} is similar to Fig. \ref{fig2a}. This shows that the kinds of nonlinear electrodynamics can not effect on the values of $\beta_{\mathrm{min}}$ and also $f$. This caused that we avoid to bring the same figures of $f$ for LN electrodynamics in Figs. \ref{figure3} and \ref{figure4}.\\
We have studied the behaviors of $f(\rho)$ versus $\rho$ for different values of $\lambda$, $\mu$ and $c$ for EN electrodynamics in Fig. \ref{figure3}. It is obvious that the value of $r_{+}$ is independent of the values of $\lambda$, $\mu$ and $c$. We can also realize this statement by extracting the constant $M$ using $f(r_{+})=0$,
\begin{eqnarray}
M=(n-1) r_{+}^{n}+\left\{
\begin{array}{ll}
$$-\frac{4 l^2\beta^2}{n} r_{+}^{n}-\frac{4(n-1)^2\beta q l^{n-1}}{n(n-2)}(\frac{l^{n-3}q}{\beta})^{\frac{1}{n-1}}(-L_{W}(-\eta_{+}))^{\frac{n-2}{2(n-1)}}\times{}_2 F_{1}([\frac{n-2}{2(n-1)}]\,,[\frac{3n-4}{2(n-1)}]\,,\\
-\frac{1}{2(n-1)}L_{W}(-\eta_{+}))+4\beta q l^{n-1} r_{+}[-L_{W}(-\eta_{+})]^{\frac{1}{2}}\times[1+\frac{1}{n}(-L_{W}(-\eta_{+}))^{-1}]$$,\quad \quad\quad\quad \ {EN}\quad &  \\ \\
$$\frac{8(2n-1)}{n^2}\beta^2 l^2r_{+}^{n}[1-\sqrt{1-\eta}]-\frac{8(n-1)^2q^2l^{2n-4}}{n^2 (n-2)r_{+}^{n-2}}{}_2 F_{1}([\frac{n-2}{2(n-1)},\frac{1}{2}]\,,[\frac{3n-4}{2(n-1)}]\,,\eta)\\-\frac{8}{n}l^2\beta^2 r_{+}^{n}\mathrm {ln}[\frac{2-2\sqrt{1-\eta}}{\eta}]$$,\quad\quad\quad\quad\quad \quad\quad\quad\quad \quad\quad\quad\quad \quad\quad\quad\quad \quad\quad\quad\quad \quad\quad\quad\quad \quad \ {LN}\quad &
\end{array}
\right.
\end{eqnarray}
where $\eta_{+}=\eta(r=0)=\frac{q^2 l^{2n-6}}{\beta^2 r_{+}^{2n-2}}$. We can find out from the above equation that the value of $r_{+}$ is not related to the values of parameters $\lambda$, $\mu$ and $c$. Also, for $\rho$ near $r_{+}$, $f$ has a similar behavior and is independent of these parameters. However, for larger $\rho$ and fixed values for parameters $q$, $M$, $n$ and $\beta$, the function $f$ depends on the values of $\lambda$, $\mu$ and $c$. In this region, by choosing small values for $\lambda$ and $c$ and a large value for $\mu$, we can have a more region of $\rho$ for which $f$ is real. For example, in Fig. \ref{fig3a}, $f$ is real for more values of $\rho$ in diagrams with $\lambda=-0.3,-0.5$ than the one with $\lambda=-0.1$. \\
In Fig. \ref{figure4}, we have compared the behavior of function $f$ for different values of dimension $n$. In this case, $f$ has a similar behavior to the one in Fig. \eqref{figure1} for $\rho$ near $r_{+}$ or larger than it. But we can see that for different values of $n$, the value of $r_{+}$ is variable and it decreases as $n$ increases. Also, for near $r_{+}$, the function $f$ increases as the value of $n$ increases. In the limit $r_{+}\rightarrow\infty$, the function $f$ goes to a constant value for each value of $r_{+}$.

\begin{figure}
\centering
\subfigure[EN]{\includegraphics[scale=0.35]{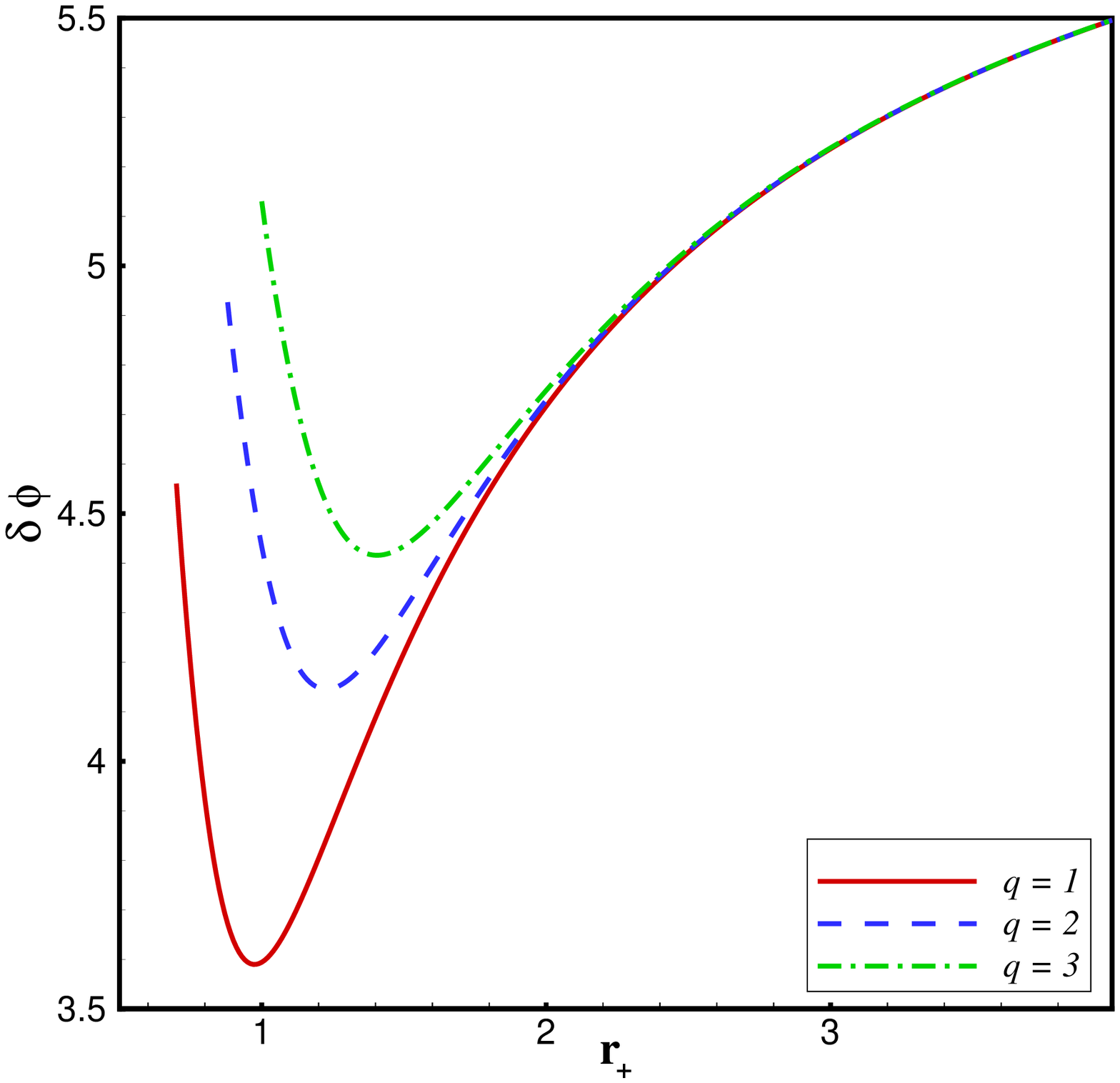}\label{fig5a}}\hspace*{.2cm}
\subfigure[LN]{\includegraphics[scale=0.35]{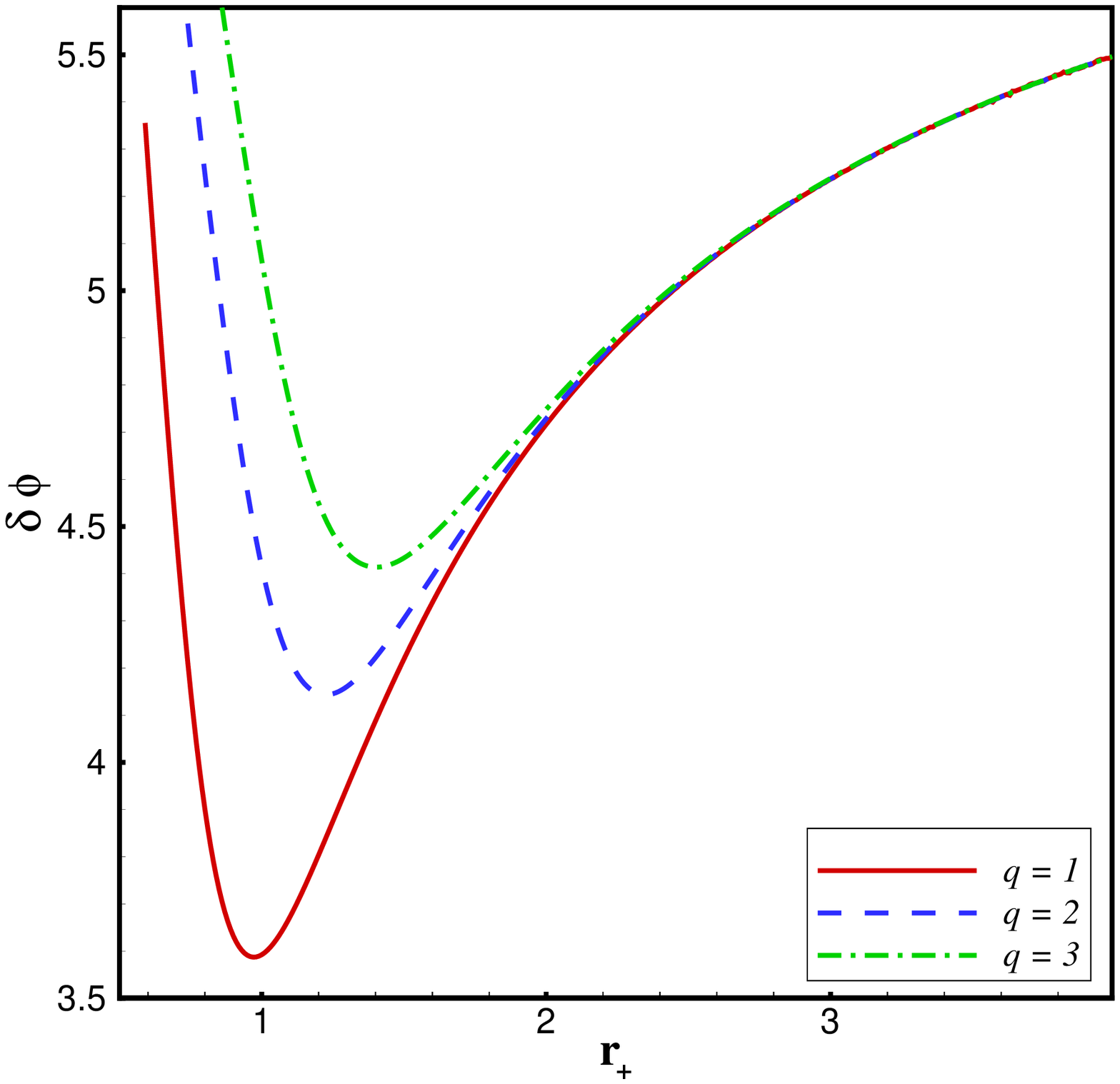}\label{fig5b}}\caption{$\delta \phi$ versus $r_{+}$ with $\beta=5$ and  $n=4$.}\label{figure5}
\end{figure}

\begin{figure}
\center
\includegraphics[scale=0.5]{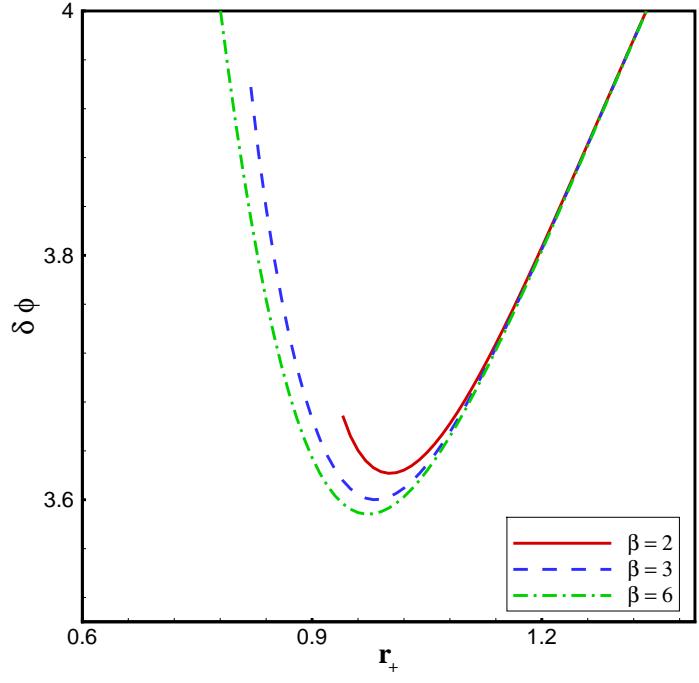}
\caption{\small{$\delta \phi$ versus $r_{+}$ with $q=1$ and  $n=4$. for EN electrodynamics.} \label{figure6}}
\end{figure}

\begin{figure}
\center
\includegraphics[scale=0.5]{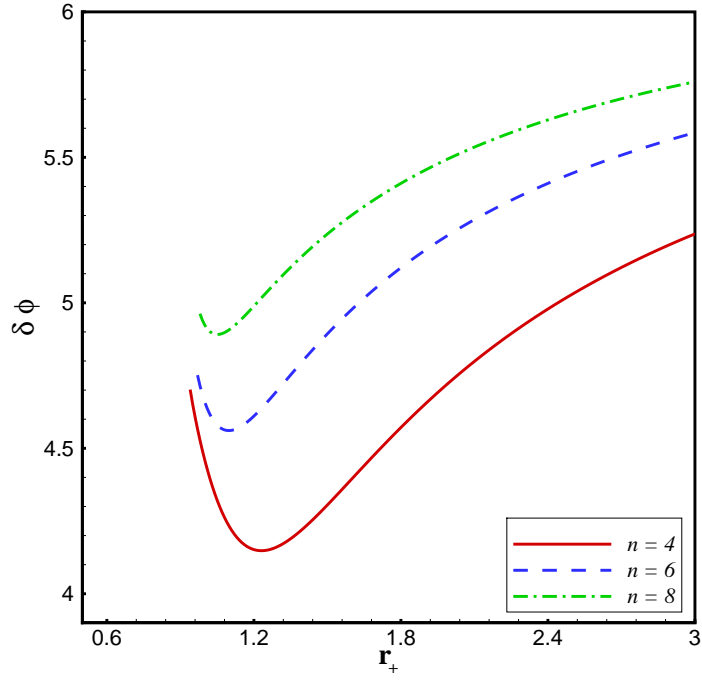}
\caption{\small{$\delta \phi$ versus $r_{+}$ with $q=2$ and  $\beta=4$. for EN electrodynamics.} \label{figure7}}
\end{figure}

Although, the Kretschmann scalar does not diverge in the range $r=[0,\infty)$, but this spacetime has a canonical singularity at $r=0$. That is, the limit of the ratio ``circumference/radius" is not $2\pi$ as the radius $r$ goes to zero. We can prove this by evaluating
\begin{eqnarray}\label{singularity}
\bigg(\mathrm{lim}_{r\rightarrow 0}\bigg(\frac{1}{r}\sqrt{\frac{g_{\phi\phi}}{g_{rr}}}\bigg)\bigg)^{-1}=\bigg(\mathrm{lim}_{r\rightarrow 0} \frac{\sqrt{r^2+r_{+}^2}l f(r)}{r^2}\bigg)^{-1}=\frac{2}{lr_{+}}\bigg(\frac{d^2 f(r)}{dr^2}\mid_{r=0}\bigg)^{-1}\ne 1,
\end{eqnarray}
where we have used of Taylor expansion for $f(r)$ at $r=0$ (or $r_{0}$)
\begin{eqnarray}
f(r)=f(r)\mid_{r_{0}}+r\frac{df(r)}{dr}\mid_{r_{0}}+\frac{r^2}{2}\frac{d^2 f(r)}{dr^2}\mid_{r_{0}}+\mathcal{O}(r^3),
\end{eqnarray}
that $f(r_{0})=\frac{df(r)}{dr}\mid_{r_{0}}=0$.
We can remove this canonical singularity at $r=0$, if we recognize the coordinate $\phi$ with the period
\begin{eqnarray}
Period_{\phi}=2\pi \bigg(\mathrm{lim}_{r\rightarrow 0}\bigg(\frac{1}{r}\sqrt{\frac{g_{\phi\phi}}{g_{rr}}}\bigg)\bigg)^{-1}=2\pi(1-4\tau),
\end{eqnarray}
where $\tau$ is obtained by equations \eqref{singularity} and \eqref{equ4}
\begin{eqnarray}\label{tau}
\tau=\frac{1}{4}\bigg[1-\frac{2l}{r_{+}^3}\bigg(\frac{d^2 \kappa}{dr^2}\mid_{r_{0}}\bigg)^{-1}\bigg].
\end{eqnarray}
So, metric \eqref{metric2} describes a locally flat spacetime that has a conical singularity at $r=0$
with a deficit angle $\delta \phi=8\pi\tau$. Now, we tend to investigate the behavior of $\delta\phi$. The first point is that according to the relation \eqref{tau}, the deficit angle parameter is
independent of the coefficients of Gauss-Bonnet and third and fourth order quasi-topological gravities and it is only dependent to the parameters $q$, $\beta$ and $n$. So, we have plotted $\delta \phi$ versus $r_{+}$ for different values of $q$, $\beta$ and $n$ in respectively three figures \eqref{figure5}, \eqref{figure6} and \eqref{figure7}. In Fig. \ref{figure5}, for each value of $q$, there is a minimum value for $r_{+}$ (we call it ${r_{+}}_{\mathrm{min}}$) that $\delta \phi$ is real only for $r_{+}>{r_{+}}_{\mathrm{min}}$. Also, there is a ${r_{+}}_{\mathrm{max}}$ that for $r_{+}>{r_{+}}_{\mathrm{max}}$, $\delta\phi$ is independent of the value $q$ and has a constant value for each value of $r_{+}$. But for ${r_{+}}_{\mathrm{min}}<r_{+}<{r_{+}}_{\mathrm{max}}$, $\delta\phi$ depends on the value of $q$ and it increases as $q$ increases. In this region, there is also a ${r_{+}}_{\mathrm{0}}$ for which $\delta\phi$ has a minimum value.\\
Although, for the same parameters $\beta$ and $n$, the values of ${r_{+}}_{\mathrm{min}}$ in LN form in Fig. \ref{fig5b} are smaller than the ones in EN form, but $\delta\phi$ has a similar behavior in both of them. So, this caused we refuse to investigate $\delta\phi$ for LN form in the next figures. \\
In Fig. \ref{figure6} and for different values of $\beta$, the general behavior of $\delta\phi$ is almost similar to the ones in Figs. \ref{fig5a} and \ref{fig5b} with a little difference. For constant values of parameters $q$ and $n$, the value of $\delta\phi$ in the region ${r_{+}}_{\mathrm{min}}<r_{+}<{r_{+}}_{\mathrm{max}}$ is related to the parameter $\beta$ and decreases as $\beta$ increases. Also, by increasing $\beta$, the value of ${r_{+}}_{\mathrm{min}}$ decreases.
Fig. \ref{figure7} has a different behavior to the two previous figures. In this figure, for each value of $r_{+}>{r_{+}}_{\mathrm{min}}$, the deficit angle increases as the dimension $n$ increases.

\section{Conserved quantities}\label{conserved}
According to our previous statements, we can not define thermodynamic quantities for magnetic branes because they are without any event horizons. Now, we would like to obtain conserved quantities of this magnetic brane such as mass density and charge. Using AdS/CFT correspondence \cite{Deh10}, we can derive the action and then the conserved quantities. For this purpose, we define the finite action
\begin{eqnarray}\label{act3}
I_{1}=I_{bulk}+I_{b},
\end{eqnarray}
where $I_{b}$ is a boundary term. $I_{b}$ makes the variational
principle well defined if we choose it as
\begin{eqnarray}
I_{b}=I_{b}^{(1)}+I_{b}^{(2)}+I_{b}^{(3)}+I_{b}^{(4)}
\end{eqnarray}
where $I_{b}^{(1)}$, $I_{b}^{(2)}$, $I_{b}^{(3)}$ and $I_{b}^{(4)}$ are respectively the proper surface terms for Hilbert-Einstein \cite{Gibb1}, Gauss-Bonnet\cite{Myer1,Deh11}, third order \cite{Deh7} and fourth order quasi-topological \cite{Deh6} gravities which are obtained as
\begin{eqnarray}
I_{b}^{(1)}=\frac{1}{8\pi}\int_{\partial\mathcal{M}} d^{n}x \sqrt{-\gamma}K,
\end{eqnarray}
\begin{eqnarray}
I_{b}^{(2)}=\frac{1}{8\pi}\int_{\partial\mathcal{M}} d^{n}x \sqrt{-\gamma}\frac{2\lambda l^2}{3(n-2)(n-3)}(3KK_{ac}K^{ac}-2K_{ac}K^{cd}K_{d}^{a}-K^3),
\end{eqnarray}
\begin{eqnarray}
I_{b}^{(3)}&=&\frac{1}{8\pi}\int_{\partial\mathcal{M}} d^{n} x \sqrt{-\gamma}\bigg\{\frac{3\mu l^4}{5n(n-2)(n-1)^2 (n-5)}(nK^5-2K^3K_{ab}K^{ab}+4(n-1)K_{ab}K^{ab}K_{cd}K^{d}_{e}K^{ec}\nonumber\\&&-(5n-6)K K_{ab}[nK^{ab}K^{cd}K_{cd}-(n-1)K^{ac}K^{bd}K_{cd}])\bigg\},
\end{eqnarray}
\begin{eqnarray}
I_{b}^{(4)}&=&\frac{1}{8\pi}\int_{\partial\mathcal{M}} d^{n} x \sqrt{-\gamma}\frac{2c l^6}{7n(n-1)(n-2)(n-7)(n^2-3n+3)}\bigg\{\alpha_{1}K^3K^{ab}K_{ac}K_{bd}K^{cd}+\alpha_{2}K^2K^{ab}K_{ab}K^{cd}K^{e}_{c}K_{de}\nonumber\\
&&+\alpha_{3}K^2K^{ab}K_{ac}K_{bd}K6{ce}K^{d}_{e}+\alpha_{4} K K^{ab}K_{ab}K^{cd}K^{e}_{c} K^{f}_{d}K_{ef}+\alpha_{5} K K^{ab} K^{c}_{a}K_{bc}K^{de}K^{f}_{d}K_{ef}+\alpha_{6} K K^{ab}K_{ac} K_{bd} \nonumber\\
&& K^{ce}K^{df}K_{ef}+\alpha_{7} K^{ab}K^{c}_{a}K_{bc}K^{de}K_{df}K_{eg}K^{fg}\bigg\}.
\end{eqnarray}
In the above terms, $\gamma_{\mu\nu}$ is the induced metric on the boundary $\partial \mathcal{M}$ and $K^{ab}$ is the extrinsic curvature of this boundary with the trace $K$.\\
The evaluated conserved quantities of the action \eqref{act3} have a problem that they are divergent. To solve this problem and define a finite action for asymptotically AdS solutions with flat boundary,
$\hat{R}_{abcd}(\gamma)=0$, we use counterterm method inspired by AdS/CFT correspondence. In this method, we add a new term $I_{ct}$ to the action \eqref{act3} to have a divergence free stress-energy tensor \cite{Henni}. $I_{ct}$ is defined as
\begin{eqnarray}
I_{ct}=-\frac{1}{8\pi}\int_{\partial\mathcal{M}} d^{n}x \sqrt{-\gamma}\frac{(n-1)}{l_{eff}},
\end{eqnarray}
where $l_{eff}$ is a scale length factor that is related to $l$ and the coefficients of gauss-Bonnet and quasi-topological gravities. It also reduces to $l$ as these coefficients go to $0$.\\
To compute the conserved quantities, we first choose a spacelike
surface $\mathcal{B}$ in $\partial \mathcal{M}$ with metric $\sigma_{ij}$ and then write the boundary metric in ADM form
\begin{eqnarray}
\gamma^{ab}dx^{a}dx^{b}=-N^2 dt^2+\sigma_{ij}(d\phi^{i}+V^{i} dt)(d\phi^{j}+V^{j} dt).
\end{eqnarray}
$N$ and $V^{i}$ are respectively the lapse and shift functions and
the coordinates $\phi^{i}$ are the angular variables parameterizing the hypersurface of constant r around the origin. If we evaluate the finite stress tensor $T_{ab}$ by the new finite action, we can obtain the quasilocal conserved quantities
\begin{eqnarray}
\mathcal{Q}(\xi)=\int_{\mathcal{B}}d^{n-1} \phi \sqrt{\sigma} T_{ab} n ^{a} \xi^{b},
\end{eqnarray}
where $n^{a}$ is the timelike unit normal vector to
the boundary $\mathcal{B}$ and $\sigma$ is the determinant of the metric $\sigma_{ij}$. $\xi^{b}$ is a Killing vector field on the boundary that we can obtain the total mass per unit volume $V_{n-1}$ by its dedicated Killing vector $\xi=\partial /\partial t$ as
\begin{eqnarray}
M_{total} =\frac{M}{4(n-1)}.
\end{eqnarray}
We can be sure that the obtained mass is finite because we have used the limit in which the boundary $\mathcal{B}$ becomes infinite. The next step is to determine the electric charge of this magnetic brane. To obtain the electric charge of the spacetimes with a longitudinal magnetic field, we should consider the projections of the electromagnetic field tensors on special hypersurfaces with normal
$u^{0} =\frac{1}{N}$, $u^{r}=0$ and $u^{i}=−\frac{N^{i}}{N}$. So, the electric field is described as
\begin{eqnarray}
E^{u}=g^{\mu\rho}F_{\rho\nu}u^{\nu}.
\end{eqnarray}
By calculating the flux of the electromagnetic field at infinity, the electric charge per unit volume $V_{n-1}$ is obtained zero. The zero value of the electric charge returns to the zero value of the electric field. Electric field is obtained when the magnetic brane has at least one rotation. As we have considered a static magnetic brane, so this brane has no electric field and followed by, no electric charge.\\
\section{concluding results}\label{result}
At last, we want to have a brief conclusion of this magnetic brane. We started our theory with an $(n+1)$-dimensional action in quartic quasi-topological gravity that is coupled to the exponential and logarithmic forms of the nonlinear electrodynamics. Quasi-topological gravity is a comprehensive higher derivative theory that leads to at most second order field equations and has no limitations on dimensions. The theory reduces to Einstein's theory, as we choose the coefficients of quasi-topological gravity zero ($\lambda=\mu=c=0$). Nonlinear electrodynamics theory is also a nonlinear theory to remove some problems such as the divergence of the electromagnetic field of Maxwell theory in the origin. This theory reduces to linear Maxwell one, as the nonlinearity parameter $\beta$ goes to infinity.\\
For our purpose, we used the metric of the spacetime that has a magnetic brane interpretation with characteristics $(g_{\rho\rho})^{-1}\propto g_{\phi\phi}$ and $g_{tt}\propto -\rho^2$.
The obtained solutions included an electromagnetic field ($F_{\phi\rho}$) that is related to the only nonzero component of the vector potential $A_{\phi}(r)$. The other solution ($f(\rho)$) was made from a fourth order field equation and was without any horizons and curvature singularities. The allowed region for $f$ is defined in the interval $r_{+}<\rho<\infty$ where does not contain the point $\rho=0$. We then investigated the behaviors of the function $f$ for different parameters. In these figures, exponential and logarithmic forms of nonlinear electrodynamics theory had the same effects on the function $f$. We also proved that the value of $r_{+}$ is independent of the values of the coefficients of Love-Lock and quasi-topological gravities ($\mu$, $\lambda$ and $c$) and showed this in the figures. $r_{+}$ increases as $q$ increases or $n$ decreases separately. For $\rho$ near $r_{+}$, the behavior of $f$ is dependent to the parameters $q$ and $n$ and independent of the values of $\lambda$, $\mu$ and $c$. In this region and for each value of $\rho$, by increasing the value of $q$ or decreasing the value of $n$ separately, the function $f$ decreases. At larger $\rho$, the function $f$ behaves vice versa and it is independent of the values of parameters $q$, $\beta$ and $n$ but is related to the parameters $\lambda$, $\mu$ and $c$. Also, for constant values of $q$, $\beta$ and $n$, the function $f$ is real for more regions of $\rho$, if we choose small values for $\lambda$ and $c$ and large value for $\mu$. There is also a $\beta_{\mathrm{min}}$ for each value $\rho$ that for $\beta>\beta_{\mathrm{min}}$, the function $f$ has a constant value. The value of $\beta_{\mathrm{min}}$ also depends on the value of $\rho$ and increases by decreasing the value of $\rho$.\\
The solutions of this magnetic brane have a conic singularity at $r=0$ with a deficit angle $\delta\phi$. We proved that the deficit angle is not related to the coefficients of Love-Lock and quasi-topological gravities and is dependent only to the parameters $q$, $\beta$ and $n$. So, we investigated the behavior of $\delta\phi$ versus $r_{+}$ for different values of $q$, $\beta$ and $n$ in some figures. There are two ${r_{+}}_{\mathrm{min}}$ and ${r_{+}}_{\mathrm{max}}$ that for ${r_{+}}_{\mathrm{min}}<r_{+}<{r_{+}}_{\mathrm{max}}$, the function $f$ is dependent to the values of $q$ and $\beta$ and it increases as $q$ increases or $\beta$ decreases separately. But for $r_{+}>{r_{+}}_{\mathrm{max}}$, the function $f$ is independent of the values of $q$ and $\beta$. For different $n$, there is no ${r_{+}}_{\mathrm{max}}$ and for all $r_{+}>{r_{+}}_{\mathrm{min}}$, the function $f$ increases, as $n$ increases. The figures also showed that although the kinds of nonlinearity can not change the general behaviors of $\delta\phi$, but they cause a few change. For example, for a constant value of $q$, They cause different values for $r_{+}$.\\
As this magnetic brane did not have any horizons, so we could not consider thermodynamics for it. We just obtained conserved quantities such as mass density and electric charge by using the counterterm method. The mass per unit volume $V_{n-1}$ has a finite value and the electric charge is zero because there is no electric field. It was clear that both of these quantities were independent of the nonlinearity parameter $\beta$.\\
In our next study, we would like to generalize this static spacetime to the case of rotating solutions with one and more rotation parameters.

\acknowledgments{We would like to thank Payame Noor University and Jahrom
University.}

\end{document}